\documentclass[aps,prl,twocolumn,superscriptaddress,showpacs,amsmath,amssymb]{revtex4}

\usepackage{graphicx}
\usepackage{bm}
\usepackage{amssymb}
\usepackage{hyperref}

\begin{document}

\title{Does a Surface Polariton Have Spin?}

\author{Konstantin Y. Bliokh}
\affiliation{Advanced Science Institute, RIKEN, Wako-shi, Saitama 351-0198, Japan}
\affiliation{A. Usikov Institute of Radiophysics and Electronics, NASU, Kharkov 61085, Ukraine}

\author{Franco Nori}
\affiliation{Advanced Science Institute, RIKEN, Wako-shi, Saitama 351-0198, Japan}
\affiliation{Physics Department, University of Michigan, Ann Arbor, Michigan 48109-1040, USA}

\begin{abstract}
We consider a $p$-polarized surface electromagnetic wave (a classical surface polariton) at the interface between the vacuum and a metal or left-handed medium. We show that the evanescent electromagnetic waves forming the surface polariton inevitably possess a backward \textit{spin energy flow}, which, together with a superluminal orbital energy flow, form the total Poynting vector. This spin energy flow generates a well-defined (but not quantized) \textit{spin angular momentum} of surface polaritons which is \textit{orthogonal} to the propagation direction. The spin of evanescent waves arises from the imaginary longitudinal component of the electric field which makes the polarization effectively \textit{elliptical} in the propagation plane. We also examine the connection between the spin and \textit{chirality} of evanescent modes.
\end{abstract}

\pacs{42.50.Tx, 42.25.Ja, 73.20.Mf}

\maketitle

\textit{Introduction.---}
The \textit{spin angular momentum} (AM) of light arises from the \textit{circular} polarization of propagating electromagnetic waves and is directed \textit{along} the wave momentum \cite{OAM}. It obviously vanishes for a linearly-polarized wave. Furthermore, the spin AM is known to be purely intrinsic, i.e., independent of the coordinate origin \cite{Berry1997,Padgett2002,Li,Bliokh}. At the same time, propagating light can also posses \textit{orbital} AM: it originates from phase gradients and can have both intrinsic and extrinsic parts \cite{OAM,Berry1997,Padgett2002,Li,Bliokh}. 
The spin and orbital AM are produced, respectively, by the local \textit{spin} and \textit{orbital energy flows} (EFs) \cite{Li,Bliokh,AP2000,BS2007,Berry2009,BBS}. These are associated with vector and scalar properties of the field and together constitute the Poynting vector (momentum density). The separation of the spin and orbital parts of the AM and Poynting vector is unique, and both parts are separately observable -- e.g., via the motion of test particles \cite{Padgett2002,Berry2009,Friese,Padgett2003,Dogariu} or the evolution of instantaneous distributions of the wave field \cite{Bekshaev}. The properties of the spin and orbital EFs in optical fields were recently examined in detail  \cite{BBS}, apart from those in \textit{evanescent waves}.

In this paper we investigate the spin and orbital properties of \textit{linearly}-polarized evanescent electromagnetic waves by considering a $p$-polarized surface polariton at the interface between the vacuum and a negative-permittivity medium \cite{Maier,Nkoma,Darmanyan,RMP}. We demonstrate that, despite its linear polarization, the evanescent wave inevitably carries non-zero \textit{spin EF} and \textit{spin AM}, the latter being directed \textit{orthogonally} to the wave momentum. Moreover, the orbital EF is \textit{super}luminal, whereas the spin EF is \textit{backward}, which together ensures \textit{sub}luminal local energy transport in the forward direction. The spin of the evanescent wave arises from the imaginary longitudinal electric field, which generates a rotation of the electric-field vector within the propagation plane. Furthermore, we examine the relations between spin and \textit{chirality} \cite{Tang,BN} for evanescent waves.

\textit{Scalar and vector features of evanescent waves.---}
We consider a $p$-polarized surface polariton plane wave at the $z=0$ interface between the vacuum ($z>0$) and a medium ($z<0$) with real permittivity $\varepsilon = \varepsilon_m < 0$ and permeability $\mu = \mu_m$. Assuming that the surface mode propagates along the $x$-axis, its unit-amplitude electric and magnetic complex fields can be written as \cite{Maier,Nkoma,Darmanyan}
\begin{eqnarray}\label{eqn:1}
{\bf E}^+\!&\!=\!&\!\left({{\bf\hat{z}} - i\frac{\kappa^+}{k_p}{\bf\hat{x}}}\right)\!f^+,~
{\bf E}^- = \varepsilon_m^{-1}\!\left({{\bf\hat{z}} + i\frac{\kappa^-}{k_p}{\bf\hat{x}}}\right)\!f^-,\nonumber\\
{\bf H}^+\!&\!=\!&\!-\frac{k_0}{k_p}\,{\bf\hat{y}}\,f^+~,~~~~~~~~{\bf H}^- = -\frac{k_0}{k_p}\,{\bf\hat{y}}\,f^-,
\end{eqnarray}
where the ``$+$'' and ``$-$'' superscripts denote quantities in the $z>0$ and $z<0$ half-spaces, and $f^\pm = \exp\left[{ik_p x \mp \kappa^\pm z - i\omega_0 t}\right]$ are the scalar wave functions localized at the interface. Here $\omega_0$ is the frequency, $k_0=\omega_0/c$, whereas the evanescent waves $f^\pm$ are characterized by complex wave vectors ${\bf k}^\pm   = k_p {\bf\hat{x}} \pm i\kappa^\pm {\bf\hat{z}}$, which satisfy the dispersion relations ${{\bf k}^\pm}^2 = k_p^2 - {\kappa^\pm}^2 = \varepsilon\mu k_0^2$. Using the proper boundary conditions at the interface, this yields the surface-polariton parameters \cite{Maier,Nkoma,Darmanyan}:
\begin{equation}\label{eqn:2}
k_p = k_0 \sqrt{\frac{{\varepsilon_m^2 - \varepsilon_m \mu_m}}{{\varepsilon_m^2 - 1}}},~
\kappa^+ = - \varepsilon_m^{-1}\kappa^- = \sqrt{k_p^2 - k_0^2}.
\end{equation}

We would like to emphasize two important features of the solutions (1) and (2). First, the surface polariton propagates along the $x$-axis with the wave number $k_p > k_0$, and its phase velocity is $v_{\rm{ph}} = c\,k_0 /k_p < c$. At the same time, the local energy-transport velocity (which in free space becomes the group velocity) can be determined using the relativistic relation between the energy $W$ and momentum $p$: $p = v_{\rm{g}} W/c^2$. For the \textit{scalar} evanescent waves $f^\pm$, the $x$-component of the momentum and energy are proportional to $k_p$ and $\omega_0$, and we arrive at the \textit{superluminal group velocity} $v_{{\rm g}O}  = c\,k_p /k_0 > c$. (In contrast, for a plane wave in free space, propagating at some angle with respect to the $x$-axis, with $k_x < k_0$, we would obtain $v_{\rm{ph}} = c\,k_0 /k_x > c$ and $v_{{\rm g}O} = c\,k_x /k_0 < c$.) Thus, it might seem that the apparent superluminal group velocity of the scalar evanescent waves $f^\pm$ contradicts relativity. 

Second, consider the polarization of the surface polariton (1). Although it can be regarded as a \textit{linearly}-polarized $p$ mode with the electric field lying in the propagation $(x,z)$ plane, we emphasize the imaginary character of the longitudinal $x$-component of the field. It arises from the transversality condition ${\bf E}^\pm \cdot {\bf k}^\pm = 0$ with imaginary $k_z^\pm = \pm i\kappa^\pm$. This results in the $\mp\pi/2$ phase difference between the $E_x^\pm$ and $E_z^\pm$ field components, i.e., in the \textit{rotation} of the electric field in the $(x,z)$ plane. In other words, a $p$-polarized evanescent wave is, in fact, \textit{elliptically} polarized in the propagation plane. 

Figure~1 shows the temporal evolution of the real electric field $\bm{\mathcal E}\left({{\bf r},t} \right) = {\rm Re} {\bf E}\left( {{\bf r},t} \right)$ and instantaneous intensity ${\mathcal I}\left( {{\bf r},t} \right) = \left| {{\rm Re} {\bf E}\left( {{\bf r},t} \right)} \right|^2  + \left| {{\rm Re} {\bf H}\left( {{\bf r},t} \right)} \right|^2$ for the surface polariton (1). The motion of the wave crests demonstrates subluminal phase velocity $v_{\rm ph}$, whereas the electric-field vector rotates in each point anticlockwise (clockwise) at $z>0$ ($z<0$). A nice interplay of these features is revealed below.
%
\begin{figure}[t]
\includegraphics[width=8.6cm, keepaspectratio]{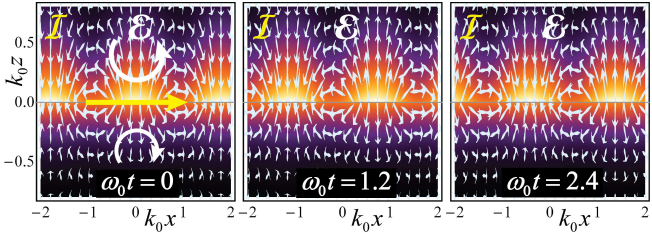}
\caption{(color online). Instantaneous distributions of the real electric field $\bm{\mathcal E}\left({{\bf r},t} \right)$ and intensity ${\mathcal I}\left({{\bf r},t} \right)$ for a surface polariton (1) propagating along the surface of a metal ($z<0$) with $\varepsilon_m = -1.5$ and $\mu_m = 1$. The electric field in each point rotates anticlockwise (clockwise) at $z>0$ ($z<0$), whereas the intensity wave crests move with the phase velocity $v_{\rm{ph}} = c\,k_0 /k_p < c$.} \label{fig1}
\end{figure}

\textit{Spin and orbital energy flows.---}
The time-averaged energy density and local EF (the Poynting vector) of an electromagnetic wave in an isotropic medium with real $\varepsilon$ and $\mu$ are given by \cite{LL}
\begin{equation}\label{eqn:3}
W = \frac{g}{2}\left[{\tilde\varepsilon \left|{\bf E}\right|^2 + \tilde\mu \left|{\bf H}\right|^2}\right],~
{\bf P} = cg\,{{\rm Re}} \left[{{\bf E}^* \times {\bf H}}\right],
\end{equation}
where $\tilde\varepsilon = d\left({\omega {\kern 1pt} \varepsilon} \right)/d\omega > 0$, $\tilde\mu = d\left({\omega {\kern 1pt} \mu} \right)/d\omega > 0$, and $g = \left({8\pi}\right)^{-1}$ in Gaussian units. The EF determines the density of the kinetic momentum of the field, ${\bf p} = {\bf P}/c^2$. Generalizing previous free-space results \cite{Li,Bliokh,AP2000,BS2007,Berry2009,BBS}, the Poynting vector in the medium can be divided into its \textit{spin} and \textit{orbital} parts, ${\bf P} = {\bf P}_S + {\bf P}_O$, as
\begin{eqnarray}\label{eqn:4-5}
{\bf P}_S = \frac{{cg}}{{4k_0}}{\rm Im} \left[{\mu^{-1} \nabla\!\times\!\left({{\bf E}^*\!\times\!{\bf E}} \right) + \varepsilon^{-1}\nabla\!\times\!\left({{\bf H}^*\!\times\!{\bf H}} \right)} \right],\\
{\bf P}_O = \frac{{cg}}{{2k_0}}{\rm Im} \left[{\mu^{-1} {\bf E}^*\!\cdot\!\left(\nabla \right){\bf E} + \varepsilon^{-1} {\bf H}^*\!\cdot\!\left(\nabla \right){\bf H}} \right].
\end{eqnarray}
The orbital EF is essentially determined by the phase gradient of the scalar wave function $f$, whereas the spin EF is produced by the gradients of the polarization ellipticities $\bm{\varphi}_E \equiv {\rm Im} \left({{\bf E}^*\times{\bf E}}\right)$ and $\bm{\varphi}_H \equiv {\rm Im} \left({{\bf H}^*\times{\bf H}}\right)$.

The separation (3)--(5) works well in a homogeneous medium, but in the presence of inhomogeneities (e.g., interfaces), the spin and orbital EFs acquire non-zero divergences: $\nabla\cdot{\bf P}_S = - \nabla\cdot{\bf P}_O \neq 0$, which does not makes a physical sense. Since $\nabla\cdot{\bf P}=0$, one can modify the separation of the spin and orbital EFs, ${\bf P} = {\bf P}^{\prime}_S + {\bf P}^{\prime}_O$, such that $\nabla\cdot{\bf P}^{\prime}_S=\nabla\cdot{\bf P}^{\prime}_O=0$ (cf. \cite{Bliokh}). In this manner, we obtain ${\bf P}^{\prime}_S={\bf P}_S+ \bm{\Delta}$, ${\bf P}^{\prime}_O={\bf P}_O - \bm{\Delta}$, with
\begin{eqnarray}\label{eqn:6}
\bm{\Delta} = \frac{cg}{4k_0} \left[\nabla\mu^{-1}\!\times\!\bm{\varphi}_E+\nabla\varepsilon^{-1}\!\times\!\bm{\varphi}_H \right].
\end{eqnarray}
This term describes a ``spin-orbit interaction'' which vanishes in a homogeneous medium, but becomes important at interfaces. 

Due to the above-mentioned polarization properties of the $p$-polarized evanescent waves, the electric-field ellipticity does not vanish for the surface polariton (1) and yields
\begin{equation}\label{eqn:7}
\bm{\varphi}_E^+ = -2\frac{\kappa^+}{k_p}{e}^{-2\kappa^+\!z}{\bf\hat{y}}~,~~
\bm{\varphi}_E^- = 2\frac{\kappa^-}{\varepsilon_m^2 k_p}{e}^{2\kappa^-\!z}{\bf\hat{y}}~.
\end{equation}
Owing to the strong $z$-gradient, this ellipticity results in a non-zero spin EF (4). Substituting Eqs. (1), (2) and (7) into Eqs. (4) and (5), we obtain
\begin{eqnarray}\label{eqn:8-9}
{\bf P}_S^+ = -\frac{cg{\kappa^+}^2}{{k_0 k_p}}{e}^{-2\kappa^+\!z}{\bf\hat{x}},~
{\bf P}_S^- = -\frac{cg{\kappa^-}^2}{{\mu_m \varepsilon_m^2 k_0 k_p}}{e}^{2\kappa^-\!z}{\bf\hat{x}},\\
{\bf P}_O^+ = \frac{cg k_p}{k_0}{e}^{-2\kappa^+\!z}{\bf\hat{x}},~~~~~
{\bf P}_O^- = \frac{cgk_p}{\mu_m \varepsilon_m^2 k_0}{e}^{2\kappa^-\!z}{\bf\hat{x}}.~~~~~
\end{eqnarray}
Accordingly, the total Poynting vector of the surface polariton is \cite{Maier,Nkoma,Darmanyan}
\begin{equation}\label{eqn:10}
{\bf P}^+ = \frac{cg k_0}{k_p}{e}^{-2\kappa^+\!z}{\bf\hat{x}}~,~~~
{\bf P}^- = \frac{cgk_0}{\varepsilon_m k_p}{e}^{2\kappa^-\!z}{\bf\hat{x}}.
\end{equation}
Importantly, because of the discontinuity of $\mu^{-1}\bm{\varphi}_E$ at the vacuum-medium interface, Eq.~(7), strong counter-propagating \textit{boundary} spin and orbital EFs arise there. Taking into account the ``spin-orbit'' correction (6), these boundary EFs are
\begin{equation}\label{eqn:11}
\delta{\bf P}_S = -\delta{\bf P}_O = 
\frac{cg \kappa^+}{2 k_0 k_p}\left(1-\frac{1}{\varepsilon_m \mu_m}\right)\delta(z).
\end{equation}
%
%
\begin{figure}[t]
\includegraphics[width=8.6cm]{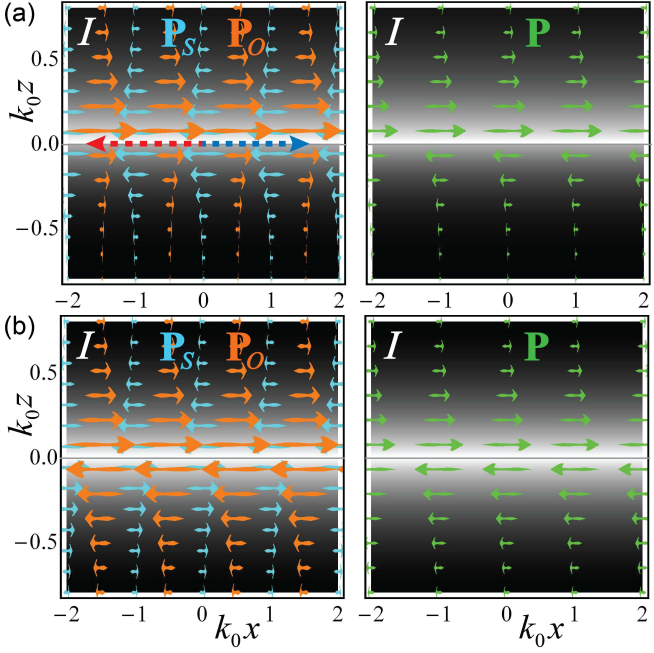}
\caption{(color online). Distributions of the time-averaged intensity $I = \left|{\bf E}\right|^2 + \left|{\bf H}\right|^2$, spin (${\bf P}_S$), orbital (${\bf P}_O$), and total (${\bf P}$) EFs (8)--(11) for the surface polariton (1) propagating along the surface of (a) a metal ($\varepsilon_m = -1.5$, $\mu_m = 1$) and (b) a ``perfect'' left-handed medium ($\varepsilon_m = \mu_m = -1$). The boundary flows (11) are indicated by dashed arrows. In the vacuum ($z>0$), the backward spin EF is subtracted from the forward superluminal orbital EF to provide subluminal energy transport. In the metal, the spin EF dominates over the orbital EF, which results in the backward energy transport \cite{Nkoma}.} \label{fig2}
\end{figure}
%

Thus, evanescent waves possess a \textit{backward spin EF} (8), ${\bf P}_S \parallel -{\bf\hat{x}}$, (in the medium it can be forward if $\mu_m<0$). This spin EF is subtracted from the forward orbital EF (9) to give the total energy current (10). Figure~2 shows the distributions of the time-averaged field intensity and EFs (8)--(11) for polaritons on the surface of (a) a metal and (b) a ``perfect'' left-handed medium with $\varepsilon_m = \mu_m = -1$ \cite{Pendry}. To understand the importance of the spin EF, note that Eqs.~(8)--(10) in the vacuum can be written as ${\bf P}_S^+ = -c\left({{\kappa^+}^2/k_0 k_p}\right)W^+ {\bf\hat{x}}$, ${\bf P}_O^+ = c\left({k_p /k_0}\right)W^+ {\bf\hat{x}}$, ${\bf P}^+ = c\left({k_0 /k_p}\right)W^+ {\bf\hat{x}}$, where $W^+ = ge^{-2\kappa^+\!z}$ is the energy density (3). Using the relation $P = pc^2 = v_{\rm g} W$, one can see that the orbital EF corresponds to the superluminal group velocity, $v_{{\rm g}O} = c\,k_p /k_0 > c$, mentioned above, while the backward spin EF reduces the total momentum and the corresponding group velocity becomes \textit{subluminal}: $v_{\rm g} = c\,k_0 /k_p = v_{\rm ph} < c$. 
Hence, it is the backward spin EF that ensures proper local energy transport in evanescent electromagnetic fields.

\textit{Angular momenta and chirality.---}
The spin and orbital parts of the wave momentum ${\bf p} = {\bf P}/c^2$ determine the spin and orbital AM of the electromagnetic field \cite{Li,Bliokh,AP2000,BS2007,Berry2009,BBS}. Their spatial densities are given by ${\bf S} = {\bf r}\times{\bf p}_S$ and ${\bf L} = {\bf r}\times{\bf p}_O$, whereas the integral (in our case -- integrated over $z$) values can be written as
\begin{eqnarray}\label{eqn:12}
\left\langle{\bf S}\right\rangle = \left\langle{{\bf r}\times{\bf p}_S}\right\rangle,~~ 
\left\langle{\bf L}\right\rangle = \left\langle{{\bf r}\times{\bf p}_O}\right\rangle. 
\end{eqnarray}
The spin represents a purely \textit{intrinsic} AM, while the orbital AM consists of \textit{extrinsic} and \textit{intrinsic} parts \cite{Berry1997,Padgett2002,Bliokh}:
\begin{eqnarray}\label{eqn:13}
\left\langle{\bf L}\right\rangle^{\rm ext} = \left\langle{\bf r} \right\rangle \times \left\langle{{\bf p}_O}\right\rangle,~~
\left\langle{\bf L}\right\rangle^{\rm int} = \left\langle{\bf L}\right\rangle - \left\langle{\bf L}\right\rangle^{\rm ext},
\end{eqnarray}
where $\left\langle{\bf r} \right\rangle$ is the centroid of the beam. 

To prove the intrinsic nature of the spin AM of surface polaritons, we calculate the integral spin EF (8) and (11). Remarkably, the positive boundary flow (11) precisely balances the negative bulk flow (8) and $\left\langle{{\bf P}_S}\right\rangle \equiv \int {{\bf P}_S}\,dz = 0$, akin to the case of propagating waves \cite{Li,Bliokh}. Thus, although the spin EF is crucial for the \textit{local} energy transport, it does not transfer energy \textit{globally}. This ensures that $\left\langle{\bf S}\right\rangle^{\rm ext} = \left\langle{\bf r} \right\rangle \times \left\langle{{\bf p}_S}\right\rangle \equiv 0$ \cite{Berry1997,Padgett2002,Li,Bliokh,AP2000}. At the same time, the global energy transport is realized by the orbital EF: $\left\langle{\bf P}_O\right\rangle = \left\langle{\bf P}\right\rangle  = \left(cgk_0/2\kappa^+ k_p\right)\left[{1 - \varepsilon_m^{-2}}\right]{\bf\hat{x}}$, and the ratio $\left\langle {\bf P} \right\rangle / \left\langle W \right\rangle$ yields the known group velocity of the surface polariton \cite{Nkoma,Darmanyan}.

The value of the AM is typically normalized by the integral energy $\left\langle W \right\rangle$ \cite{OAM}. Since $\left\langle W \right\rangle$ is strongly dependent on the dispersion in the medium, we first calculate the spin and orbital AM for the free-space evanescent field in the $z>0$ half-space. Using Eqs.~(1)--(3) and (8)--(10), its energy is $\left\langle {W^+} \right\rangle \equiv \int\limits_{z > 0} W dz = g/2\kappa^+$, whereas the spin and orbital AM (12) become
\begin{equation}\label{eqn:14}
\left\langle{{\bf S}^+}\right\rangle = -\frac{{\kappa^+}}{{2\omega_0 k_p}}\left\langle{W^+}\right\rangle{\bf\hat{y}},~ 
\left\langle{{\bf L}^+}\right\rangle = \frac{{k_p}}{{2\omega_0 \kappa^+}}\left\langle{W^+}\right\rangle{\bf\hat{y}}.
\end{equation}
Noteworthily, the same spin AM (in units of $\hbar$ per particle) can be obtained via calculating the normalized expectation value of the quantum spin-1 operator ${\bf\hat{S}}$ with the fields (1) \cite{Berry2009}: 
\begin{equation}\label{eqn:15}
\frac{\langle{\bf E}^+,{\bf H}^+|\,{\bf\hat{S}}\,|{\bf E}^+,{\bf H}^+\rangle}{\left\langle{\bf E}^+,{\bf H}^+\left.\right|{\bf E}^+,{\bf H}^+\right\rangle} = -\frac{\kappa^+}{2k_p}{\bf\hat{y}}.
\end{equation}
In the whole space, the AM yield
\begin{equation}\label{eqn:16}
\left\langle{{\bf S}}\right\rangle = \left(1-\frac{1}{\varepsilon^2_m \mu_m}\right)\!\left\langle{{\bf S}^+}\right\rangle,~ 
\left\langle{{\bf L}}\right\rangle = \left(1-\frac{1}{\varepsilon^2_m \mu_m}\right)\!\left\langle{{\bf L}^+}\right\rangle.
\end{equation}
Thus, evanescent waves and surface polaritons posses well-defined (but not quantized) spin and orbital AM directed \textit{orthogonally} to the propagation $(x,z)$ plane.

The separation of the intrinsic and extrinsic parts of the orbital AM is determined by the centroid of the field, $\left\langle{z}\right\rangle = \left\langle{z\,W}\right\rangle /\left\langle{W}\right\rangle$, which depends on the medium dispersion. As an example we consider polaritons on the surface of a ``perfect'' left-handed material with $\varepsilon_m \left({\omega_0}\right) = \mu_m \left({\omega_0}\right) = -1$ \cite{Pendry}. In this case, $\kappa^+ = \kappa^-$, the boundary EFs (11) vanish, and the EFs are mirror \textit{anti-symmetric} with respect to the $z=0$ plane, Fig.~2b. Choosing the model dispersions $\varepsilon_m \left(\omega\right) = \mu_m \left(\omega\right) = 1 - 2\omega_0 /\omega$, we have $\tilde\varepsilon_m \left({\omega_0}\right) = \tilde\mu_m \left({\omega_0}\right) = 1$, and the energy densities become mirror \textit{symmetric} with respect to the $z=0$ plane. In this case $\left\langle z \right\rangle = 0$, and purely intrinsic spin and orbital AM (16) yield $\left\langle{{\bf S}}\right\rangle = 2 \left\langle{{\bf S}^+}\right\rangle$ and $\left\langle{{\bf L}}\right\rangle = 2 \left\langle{{\bf L}^+}\right\rangle$.

An intrinsic AM can be associated with a circulating EF, i.e., a \textit{vortex} \cite{OAM,AP2000,BBS}. One can see that the energy circulation in the surface-polariton field is non-zero for any contour encircling the origin (see Fig.~2). In such circumstances, the vortex EF appears upon the longitudinal $x$-localization of the field, i.e., considering a surface-polariton \textit{wave packet} \cite{Shadrivov}. As an example, we consider a realistic left-handed material with plasma dispersions $\varepsilon_m \left(\omega\right) = 1 - \omega_{\varepsilon}^2 /\omega^2$, and $\mu_m \left(\omega\right) = 1 - \omega_{\mu}^2 /\omega^2$. For this case, Figure 3 shows a surface-polariton wave packet calculated numerically using a narrow Gaussian spectrum of solutions (1) centered around $\omega=\omega_0$. It is clearly seen that the spin and orbital EFs form counter-circulating vortices in the $(x,z)$ plane. In this generic case, the orbital AM (12) and (13) contains both intrinsic and extrinsic parts.
%
\begin{figure}[t]
\includegraphics[width=8.6cm, keepaspectratio]{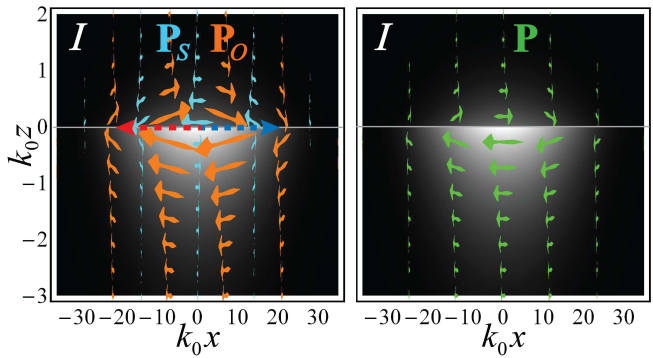}
\caption{(color online). The same intensity and EF distributions as in Fig. 2, but for a surface-polariton wave packet propagating along the surface of a dispersive left-handed material with electric and magnetic plasma frequencies $\omega_{\varepsilon} = \sqrt{3/2}\,\omega_0$ and $\omega_{\mu} = 2\,\omega_0$, which correspond to $\varepsilon_m \left({\omega_0}\right) = -0.5$ and $\mu_m \left({\omega_0}\right) = -3$. The counter-propagating EFs in the vacuum ($z>0$) and the medium ($z<0$) bring about the vortex EFs in the localized wave-packet solutions. These solutions carry spin, intrinsic orbital, and extrinsic orbital AM along the $y$-axis.} \label{fig3}
\end{figure}
%

Finally, we explore an important connection between the spin and \textit{chirality} of the wave \cite{Tang,BN}. Akin to the energy density $W$ and energy flow ${\bf P}$, one can characterize the chirality of the electromagnetic field by the \textit{chirality density} ${\rm K}$ and \textit{chirality flow} $\bm{\Phi}$ which satisfy the continuity equation \cite{Tang,BN}. Generalizing the earlier free-space results to the case of a homogeneous medium, these time-averaged quantities can be written as
\begin{equation}\label{eqn:17}
{\rm K} = g\,{\rm Im}\!\left({{\bf H}^*\!\cdot\!{\bf E}}\right),~
\bm{\Phi} = \frac{{cg}}{2}{\rm Im}\!\left({\mu^{-1}{\bf E}^*\!\times\!{\bf E} + \varepsilon^{-1}{\bf H}^*\!\times\!{\bf H}}\right).
\end{equation}
Substituting here the surface-polariton field (1), we immediately arrive at
\begin{equation}\label{eqn:18}
{\rm K} = 0,~\bm{\Phi} = \frac{{cg}}{2}\mu^{-1}\bm{\varphi}_E \ne 0.
\end{equation}
Thus, the chirality density vanishes since ${\bf H}^*\cdot{\bf E} = 0$, whereas a non-zero chirality flow is determined by the ellipticity of the field polarization, Eq.~(7). To understand these results, note that for the propagating fields, the integral chirality $\left\langle {\rm K} \right\rangle$ is intimately related to the averaged \textit{helicity} of photons, whereas the chirality momentum $\left\langle {\bm{\Phi}/c^2} \right\rangle$ is proportional to the spin AM $\left\langle {\bf S} \right\rangle$ \cite{BN}. In our case, the helicity vanishes identically because the spin AM is \textit{orthogonal} to the momentum, and $\left\langle {\rm K} \right\rangle = 0$. At the same time, calculating the integral chirality momentum, we find that it is indeed proportional to the spin AM:
\begin{equation}\label{eqn:18}
\left\langle {\bm{\Phi}/c^2 } \right\rangle = 2k_0 \left\langle {{\bf S}} \right\rangle.
\end{equation}
Here we obtained an additional factor of 2 as compared to the general result for propagating fields \cite{BN}. Noteworthily, the connection between the chirality and helicity (rather than spin) is quite fundamental. The main point is that chirality is a parity-odd but time-even property \cite{Barron}. Spin changes its sign upon time inversion, while helicity does not.

\textit{Discussion.---}
We have shown that evanescent waves in free space and surface polariton modes at the interface with a negative-permittivity medium possess superluminal orbital energy flow and non-zero backward spin energy flow. 
The latter originates from the rotation of the electric field in the plane of propagation, and it is necessary for proper energy transport. 
The EFs generate well-defined spin and orbital angular momenta which are orthogonal to the propagation direction of the wave. The helicity and chirality density naturally vanish in such case.
It is worth noticing that the previously considered AM of propagating waves \cite{OAM} and near-field vortices \cite{SP vortex} essentially require the superposition of \textit{multiple} plane waves which produce the necessary gradients. In sharp contrast, the transverse spin and orbital AM already appear here for a \textit{single} surface-polariton plane wave (two evanescent waves) owing to its natural confinement (inhomogeneity) across the interface.

Our results appeal to experimental tests revealing the unusual transverse spin of surface polariton evanescent waves. Typically, spin AM manifests itself in interactions with probe particles, and it is important to discuss the fundamentals of such experiments.

The \textit{spin AM} is usually observed in propagating optical fields via the \textit{spinning motion} of the absorbing or birefringent test particles \cite{Padgett2002,Friese,Padgett2003,Dogariu}. Any local perturbation of the field with a non-zero ellipticity (e.g., a small region of the field exclusion around the particle) immediately induces radial intensity gradients and circulating spin EF \cite{AP2000}. This circulating EF spins the particle in any point of the elliptically-polarized field. In the case of evanescent waves, the situation becomes more complicated, because any local perturbation in the $(x,z)$ plane will \textit{drift} along the $x$-axis with velocity $v_{\rm g} = v_{\rm ph} = c\,k_0 /k_p$. Still, the probing particle can experience a non-zero circulation of the spin EF which will induce its spinning motion. 

In addition to the spinning motion, test particles can move \textit{linearly} in the background EFs. Such interaction crucially depends on the physical properties of the particle. For instance, Berry noticed \cite{Berry2009} that the forces acting on small \textit{absorbing} and \textit{conducting} particles are proportional to the \textit{orbital} EF (8) and the \textit{total} Poynting vector (10), respectively. Hence, such particles in the evanescent field will experience forces $k_p/k_0$ times \textit{higher} and $k_0/k_p$ times \textit{weaker} than the analogous force from a propagating plane wave without spin EF. Thus, monitoring the linear motion of different particles, one could observe the action of different EFs.

Finally, the vanishing of the chirality density implies that the interaction of the surface-polariton plane waves with small \textit{chiral} particles (e.g., molecules) cannot distinguish between right- and left-handed enantiomers. The verification of this conclusion could also be an important confirmation of the above theory.

We acknowledge valuable discussions with A. Y. Bekshaev, Y. P. Bliokh, Y. Gorodetski, and support from the European Commission (Marie Curie Action), ARO, JSPS-RFBR contract No. 12-02-92100, Grant-in-Aid for Scientific Research (S), MEXT Kakenhi on Quantum Cybernetics, and the JSPS through its FIRST program.


\end{document}